\documentclass[final,5p,times]{elsarticle}

\usepackage{lineno,hyperref,amssymb, bm, amsmath,multirow}
\usepackage{tabulary}
\usepackage{algorithm}
\usepackage[noend]{algpseudocode}
\usepackage{xcolor}
\usepackage{tikz}

\algnewcommand{\algorithmicand}{\textbf{ and }}
\algnewcommand{\algorithmicor}{\textbf{ or }}

\algnewcommand{\OR}{\algorithmicor}
\algnewcommand{\AND}{\algorithmicand}

\algnewcommand{\LeftComment}[1]{\Statex \(\triangleright\) #1}

\newcommand{\stkout}[1]{\ifmmode\text{\sout{\ensuremath{#1}}}\else\sout{#1}\fi}

\newcommand\copyrightnotice[1]{
    \begin{tikzpicture}[remember picture,overlay]
    \setlength{\fboxsep}{0pt}
    \setlength{\fboxrule}{0pt}
    \node[anchor=south,yshift=-2pt] at (current page.south) {\fbox{\parbox{\dimexpr\textwidth-\fboxsep-\fboxrule\relax}{#1}}};
    \end{tikzpicture}
}

\journal{Acta Astronautica}

\begin{document}

\begin{frontmatter}

\title{Observation of atmospheric gravity waves using a Raspberry Pi camera module on board the International Space Station}

\date{December 29, 2020}


\author[fcul,ia]{Tiago E. C. Magalhães\corref{mycorrespondingauthor}}
\cortext[mycorrespondingauthor]{Corresponding author}
\ead{tiagoemagalhaes@fc.up.pt}

\author[escola]{Diogo E. C. G. Silva\fnref{chemistry}}
\fntext[chemistry]{Current address: Departamento de Química e Bioquímica, Faculdade de Ciências, Universidade do Porto, Rua do Campo Alegre s/n, 4169-007 Porto, Portugal}


\author[escola]{Carlos E. C. G. Silva\fnref{viana}}
\fntext[viana]{Current address: Escola Superior de Tecnologia e Gestão, Instituto Politécnico de Viana do Castelo, Avenida do Atlântico, 4900‐348 Viana do Castelo, Portugal}

\author[escola]{Afonso A. Dinis\fnref{ess}}
\fntext[ess]{Current address: Escola Superior de Saúde do Porto do Instituto Politécnico do Porto, Portugal, Rua Dr. António Bernardino de Almeida, 400, 4200-072 Porto, Portugal}

\author[escola]{José P. M. Magalhães\fnref{isep}}
\fntext[isep]{Current address: Instituto Superior de Engenharia do Porto, Instituto Politécnico do Porto, 4200‐072 Porto, Portugal}

\author[ifimup]{Tânia M. Ribeiro}

\address[fcul]{Departamento de Física, Faculdade de Ciências, Universidade de Lisboa, Edifício C8, Campo Grande, PT1749-016 Lisboa, Portugal}

\address[ia]{Instituto de Astrof{\'i}sica e Ci{\^e}ncias do Espa\c{c}o, Edif{\'i}cio C8, Campo Grande, PT1749-016 Lisboa, Portugal}

\address[escola]{Escola Secundária da Maia, Av. Luís de Camões, 4470-194 Maia, Portugal}

\address[ifimup]{Instituto de Física de Materiais Avançados, Nanotecnologia e Fotónica and Departamento de Física e Astronomia, Faculdade de Ciências, Universidade do Porto, 4169-007 Porto, Portugal}

\begin{abstract}
We identified and computed the horizontal wavelengths of atmospheric gravity waves in clouds using a visible camera installed on a window of the Columbus module of the International Space Station (ISS) and controlled by a Raspberry Pi computer. The experiment was designed in the context of the Astro Pi challenge, a project run by ESA in collaboration with the Raspberry Pi Foundation, where students are allowed the opportunity to write a code to be executed at the ISS. A code was developed to maximize the probability of capturing images of clouds while the ISS is orbiting the Earth. Several constraints had to be fulfilled such as the experiment duration limit (3 hours) and the maximum data size (3 gigabytes). After receiving the data from the ISS, small-scale gravity waves were observed in different regions in the northern hemisphere with horizontal wavelengths in the range of $1.0$ to $4.7\,\mathrm{km}$.

\end{abstract}


\end{frontmatter}

\copyrightnotice{\footnotesize © 2021. This manuscript version is made available under the CC-BY-NC-ND 4.0 license \url{http://creativecommons.org/licenses/by-nc-nd/4.0/}}


\section{Introduction}

\subsection{Background}
Gravity waves, or buoyancy waves, are disturbances in a fluid medium, or at the interface between two media, whose restoring force is gravity (through buoyancy)~\cite{holton2004}. In a stably stratified fluid, where the fluid's density increases with depth, gravity waves are able to propagate within the fluid (vertically and horizontally) and are usually referred to as internal gravity waves~\cite{holton2004,nappo2013}. For instance, in an unstably stratified fluid (e.g., boiling water), internal gravity waves are not supported~\cite{sutherland2010}. Moreover, these waves are transverse since the particles of the fluid oscillate perpendicularly to the direction of propagation. When the fluid is a planetary stably stratified atmosphere, internal gravity waves are usually designated as atmospheric gravity waves~\cite{barrow2012,vachon1995ers}. With the exception of the planetary boundary layer, the Earth's atmosphere is, in general, stably stratified~\cite{nappo2013} and, therefore, we will use the term atmospheric gravity waves for internal gravity waves in the Earth's atmosphere.

Atmospheric gravity waves are often generated in the troposphere when parcels of air are forced upward due to topography (e.g., airflow over mountains), convection, and
wind shear~\cite{fritts2003automatic}, but can also be generated in the stratosphere~\cite{yoshiki2000}. We can divide them into two groups: orographic gravity waves (i.e., induced by topography) and non-orographic gravity waves. The former ones are usually known as mountain or \textit{lee} waves. Throughout their propagation, they transport energy and momentum to the upper layers of the atmosphere~\cite{fritts2003automatic,andrews2010,lai2019}. Atmospheric gravity waves can be detected by visual patterns created in the atmosphere, for instance, in  meteor trails, airglow, and clouds. The observation techniques used to study these waves can involve remote sensing techniques or $\textit{in situ}$ observations, such as ground-based lidars, airglow images, noctilucent cloud images, radio\-sonde soundings, and satellite images~\cite{ehard2016}.

In this paper, we present results for the detection and horizontal wavelength computation of atmospheric gravity waves using cloud pictures acquired with a visible camera installed on a window of the Columbus module of the International Space Station (ISS) and controlled by a Raspberry Pi computer. We discuss the algorithm used to take pictures of the Earth during the experiment. The work was done in the context of the Astro Pi challenge, which we will next introduce.

\subsection{The Astro Pi challenge}

The Astro Pi challenge, or just Astro Pi, is an education project of the European Space Agency (ESA) in collaboration with the Raspberry Pi Foundation~\cite{honess2017}. It offers primary and secondary school students the opportunity to design an experiment and write computer codes in the Python programming language to be executed in a Raspberry Pi at the ISS. Astro Pi is divided into two different missions, a non-competitive one called Mission Zero and a competitive one called Mission Space Lab. The former one is aimed at students who are under 15 years of age (team from 2 to 4 members), while the latter is for students under 20 years old (teams from 2 to 6 members). Both missions require the supervision of a mentor (or teacher). 

The work presented in this paper is inserted in the Astro Pi Mission Space Lab. In this mission, two identical Raspberry Pi computers are located in the ISS's Columbus module that can be used for different scientific purposes, one for "life in space" and the other for "life on Earth". The main difference between them is that one (nicknamed Ed) has an infrared camera (Raspberry Pi NoIR Camera), while the other one (nicknamed Izzy) has a standard visible spectrum camera. The Izzy's camera is placed at a window pointed to Earth and, therefore, it is used to study "life on Earth". Additionally, a blue filter is coupled to Izzy's camera that blocks most of the green and red light from entering the camera.

The Cloud4 team, made up of the four high-school students co-authors of this paper, submitted an idea for this challenge, which embodied four distinct phases. The first phase comprised the design and submission of an idea for an experiment. After being selected, the team enrolled in the second phase, which consisted of the development of a code to be executed in the Raspberry Pi "Izzy" computer on board the ISS. After an evaluation procedure, the team advanced to the third phase, where the code was deployed and executed on the ISS. Finally, in the fourth phase, the data was received and analyzed.

\subsection{Outline}
This paper is organized as follows. In Section~\ref{sec:Model}, we will describe a simple static model for gravity waves, where we will define the parameters of interest that we proposed to measure in the experiment. In Section~\ref{sec:Methodology}, we will present and discuss the code we developed that was executed in a Raspberry Pi computer at the ISS to capture images of clouds on Earth. In Section~\ref{sec:Results} we will present the results we obtained, namely, the performance of the code and the detection and characterization of gravity waves. Finally, in Section~\ref{sec:Conclusion}, we draw conclusions from this work.

\section{Theoretical background}
\label{sec:Model}

\subsection{Two-dimensional static gravity wave model}
Let us consider a simple model for the propagation of an atmospheric gravity wave. We can describe its field as
\begin{equation}
    U(\mathbf{r},t)=A\cos\left(\mathbf{k}\cdot\mathbf{r}-\omega t\right)\,,\label{eq:general}
\end{equation}
where $\mathbf{r}=(x,y,z)$ is the position vector, $\mathbf{k}=(k_x,k_y,k_z)$ is the wave vector, $\omega=2\pi/T$ is the angular frequency, where $T$ is the period, $t$ is time, and $A$ is a positive constant. If $k_x$, $k_y$, and $k_z$ are real values, the wave is transporting energy both horizontally and vertically and, therefore, it is designated as an internal gravity wave~\cite{nappo2013}. Henceforth,  we will assume that the wavenumbers are real such that
\begin{equation}
    k_{m} = \frac{2\pi}{\lambda_{m}}\,\,\,\,(m=x,y,z)\,,\label{eq:wavenumber}
\end{equation}
where $\lambda_{m}$ is the wavelength in the $m$ direction. Let $\phi(z,t)$ be a variable defined as
\begin{equation}
    \phi(z,t)=\frac{k_z\,z-\omega t}{2\pi}\,.\label{eq:phi}
\end{equation}
If we substitute equations~(\ref{eq:wavenumber}) and~(\ref{eq:phi}) into~(\ref{eq:general}), we obtain
\begin{equation}
    U(\mathbf{r},t)=A\cos \left\{ 2\pi\left[\frac{x}{\lambda_x} + \frac{y}{\lambda_y} +\phi(z,t) \right] \right\} \,.\label{eq:U(r,t)2}
\end{equation}
For a time instant $t=t_0$ and a constant altitude $z=z_0$, we can set $\phi(z_0,t_0)=\phi_0$ as a constant and equation~(\ref{eq:U(r,t)2}) becomes only dependent on $x$ and $y$, i.e.,
\begin{equation}
    U(x,y)=A\cos \left[ 2\pi\left(\frac{x}{\lambda_x} + \frac{y}{\lambda_y} +\phi_0 \right) \right] \,.\label{eq:Ulast}
\end{equation}
Equation~(\ref{eq:Ulast}) represents a static two-dimensional model which can describe, for instance, gravity waves present in clouds at a given altitude and instant of time. We will use it to model the observed gravity waves in clouds. 

The horizontal wavelength defines the two-dimensional distance of one complete gravity wave cycle and is given by
\begin{equation}
    \lambda_h = \frac{\lambda_x\,\lambda_y}{\sqrt{\lambda_x^2+\lambda_y^2}}\,.\label{eq:hw}
\end{equation}
This quantity can be used to classify gravity waves. Small-scale gravity waves are defined as having a horizontal wavelength $\lambda_h<100\,\mathrm{km}$ and a period  $T<45\,\mathrm{min}$~\cite{taylor2009,essien2018}. On the other hand, medium-scale gravity waves go from a hundred kilometers and their period can be of several hours~\cite{taylor2009,essien2018}. Since equation~(\ref{eq:Ulast}) does not contain information about the period, we will only distinguish between these two types of gravity waves by analyzing their horizontal wavelength.

\begin{table*}[t]
\centering
\caption{Clouds genera description.}
\label{tab:genera}
\begin{tabular}{|>{\centering}p{0.20\linewidth}|c|c|c|>{\centering}p{0.42\linewidth}|}
\hline 
Dimensions & \multicolumn{2}{c|}{Altitude (km)} & Genera & Description\tabularnewline
\hline 
\hline 
\multirow{7}{0.90\linewidth}{Horizontal dimensions larger than the vertical dimensions} & \multirow{2}{*}{Low} & \multirow{2}{*}{$0-2$} & \textit{Stratocumulus} & Gray and white; Non-uniform layer composed of individual clouds with irregular shapes; Occasionally organized in lines.\tabularnewline
\cline{4-5} \cline{5-5} 
 &  &  & \textit{Stratus } & Gray; Continuous, featureless, and uniform gray cloud layer; 
 Usually overcast.\tabularnewline
\cline{2-5} \cline{3-5} \cline{4-5} \cline{5-5} 
 & \multirow{2}{*}{Mid} & \multirow{2}{*}{$2-7$} & \textit{Altocumulus} & White; Layer of clouds that forms a uniform pattern; Usually, a mosaic-like layer of rounded small clouds or rolls, separated by orderly clear spaces.\tabularnewline
\cline{4-5} \cline{5-5} 
 &  &  & \textit{Altostratus} & Gray; Uniform or sometimes striated cloud layer; Nearly full-sky cover. \tabularnewline
\cline{2-5} \cline{3-5} \cline{4-5} \cline{5-5} 
 & \multirow{3}{*}{High} & \multirow{3}{*}{$7-15$} & \textit{Cirrus } & White; Thin clouds, assuming the shape of straight or curved filaments.\tabularnewline
\cline{4-5} \cline{5-5} 
 &  &  & \textit{Cirrocumulus} & White; Composed of very small elements, usually tiny cloud patches or thin white ripples. \tabularnewline
\cline{4-5} \cline{5-5} 
 &  &  & \textit{Cirrostratus} & White; Uniform or fibrous cloud veil; Offers more coverage of the sky
than cirrus clouds.\tabularnewline
\hline 
\multirow{3}{0.90\linewidth}{Vertical dimensions comparable to or larger than the horizontal dimensions } & Low & $0-2$ & \textit{Cumulus} & Gray or white; Puffy, voluminous cloud with clearly defined edges. \tabularnewline
\cline{2-5} \cline{3-5} \cline{4-5} \cline{5-5} 
 & Low/Mid & $0-7$ & \textit{Nimbostratus} & Dark gray; Thick, rain- or snow-producing cloud.\tabularnewline
\cline{2-5} \cline{3-5} \cline{4-5} \cline{5-5} 
 & Low/High & $0-15$ & \textit{Cumulonimbus} & Dark gray; Dense rain-producing cloud; The largest type of clouds.  \tabularnewline
\hline 
\end{tabular}
\end{table*}

\subsection{Clouds' spatial resolution}
\label{sec:spatial_resolution}

\begin{figure}[!b]
\centering
\includegraphics[width=0.35\textwidth]{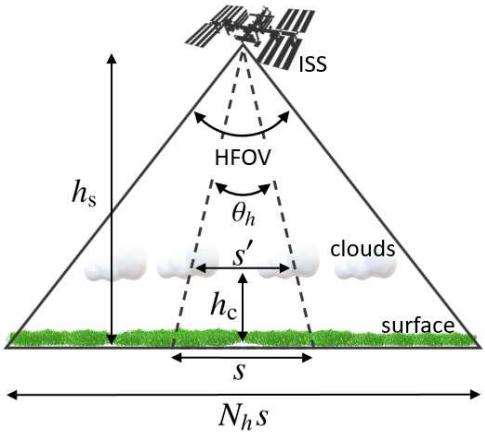}
\caption{\label{fig:height} (Color online) Notation used for the estimation of the cloud's spatial resolution $s'$ (not to scale). The dashed line comprises the instantaneous horizontal FOV, $\theta_h$, $s$ is the spatial resolution in the ground, $h_\mathrm{s}$ is the ISS altitude and $N_h$ is the number of horizontal pixels of the camera's sensor.}
\end{figure}

We will use pictures taken on board the ISS to detect gravity waves in clouds. To spatially characterize them in terms of their horizontal wavelength, we will need to estimate the spatial resolution $s'$ of clouds in a given picture. For instance, consider a satellite picture of a region on Earth containing land and clouds. The spatial resolution $s$ refers to the ground spatial resolution (Earth's surface) and not the clouds (see Fig.~\ref{fig:height}).

With reference to Fig.~\ref{fig:height}, let $h_{\mathrm{s}}$ be the altitude of the ISS and $\theta_h$ the horizontal instantaneous field of view (FOV), given by
\begin{equation}
    \theta_h = 2\,\mathrm{arctan}\left(\frac{s}{2h_{\text{s}}}\right)\,\label{eq:theta}.
\end{equation}
If clouds are present in the image, their spatial resolution will be different since their altitude is also different. Thus, the spatial resolution $s'$ of the image for clouds can be written as
\begin{equation}
    s' = 2\left(h_{\text{s}}-h_{\text{c}} \right)\mathrm{tan}\left(\frac{\theta}{2}\right)\,\label{eq:sl}.
\end{equation}
where $h_{\mathrm{c}}$ is the altitude of the clouds subject to analysis in the image. Substituting equation~(\ref{eq:theta}) into~(\ref{eq:sl}) we obtain
\begin{equation}
    s' = s \left(1-\frac{h_\mathrm{c}}{h_\mathrm{s}} \right)\,.\label{eq:SRC}
\end{equation}
Notice that if $h_\mathrm{c}=0$, then $s'=s$, as expected. It must be noted that equations~(\ref{eq:theta}) and~(\ref{eq:sl}) are valid approximations under the assumption that the spatial dimensions under analysis (i.e., $N_{h}\,s$) are very small compared to the curvature of the Earth. As we will see in Section~\ref{subsec:res_s}, this will be the case.

In summary, to determine the spatial resolution $s'$ of clouds in a given picture, three quantities are needed: (i) the ISS altitude $h_{\mathrm{s}}$, (ii) the clouds' altitude $h_{\mathrm{c}}$, and (iii) the spatial resolution $s$ of the Earth's surface. For the altitude of the ISS one can use the average value $h_\mathrm{s}=415\,\mathrm{km}$~\cite{dachev2017}. For the remaining two quantities, we will next describe how we estimated their values.

\subsubsection{Clouds' altitude}
The clouds' altitude $h_\mathrm{c}$ is estimated by identifying the type of clouds in order to find the interval of altitudes where they are typically present. Then, the average value of that interval is used as an estimation of $h_\mathrm{c}$. The altitude of a given cloud can be roughly estimated through its type, also called cloud genera. By analyzing the visual appearance of the cloud, namely, its dimension, shape, and color, we can find the altitude interval where the cloud is present~\cite{lamb2011physics}. Although this is far from an accurate method, it is particularly useful when there is no accurate information about the cloud's surrounding environment. When a mixture of more than one type of cloud is present in the same image, the evaluation of their genera and consequently the altitude may turn out to be complicated. Algorithms have been developed and optimized to help with the process of classification~\cite{heinle2010automatic}. Clouds genera are identified in Table~\ref{tab:genera} along with the atmospheric altitude level \cite{lamb2011physics, heinle2010automatic}. Low-level clouds occur up to 2 km above the surface, mid-level clouds between 2 to 7 km, and high-level clouds at 7 km or above. In this paper, we did not implement any particular algorithm for the identification of the clouds. They were classified following Table~\ref{tab:genera}.

\subsubsection{Ground spatial resolution}
\label{subsec:s}
To determine $s$, one can use two methods. The first method consists of using Google Earth to estimate the spatial resolution $s$ using land features present near the regions that are being analyzed. Hence, if $N_s$ is a distance between two land features in the image in units of pixels and $s_{\mathrm{G}}$ the distance of the same features obtained using Google Earth, the spatial resolution is estimated as
\begin{equation}
    s = \frac{s_{\mathrm{G}}}{N_s}\,.
\end{equation}  
The second method consists of using the FOV of the camera. Let $N_h$ and $N_v$ be the number of pixels in the horizontal and vertical directions, respectively, of the Raspberry Pi camera sensor. The horizontal and vertical FOV, $\mathrm{FOV}_h$ and $\mathrm{FOV}_v$, are given by
\begin{equation}
    \mathrm{FOV}_{h,v} = 2\arctan \left( \frac{N_{h,v}\,p}{2 f} \right)\,, \label{eq:fov}
\end{equation}
where $f$ is the focal length of the camera and $p$ the pixel size. Applying simple trigonometry (see Fig.~\ref{fig:height}), the spatial resolution $s$ can be calculated as
\begin{equation}
    s = \frac{2 h_\mathrm{s}}{N_{h,v}}\tan\left(\frac{\mathrm{FOV}_{h,v}}{2} \right)\,.\label{eq:s1}
\end{equation}
Substituting equation~(\ref{eq:fov}) into~(\ref{eq:s1}), we obtain
\begin{equation}
    s = \frac{h_\mathrm{s}\,p}{f}\,.\label{eq:s2}
\end{equation}
Therefore, by simply knowing the ISS altitude, the pixel size of the sensor and the focal distance, we can estimate the spatial resolution $s$.

\section{Methodology}
\label{sec:Methodology}

\subsection{Overview}
\label{subsec:overview}
The experiments in the Astro Pi Mission Space Lab have a limited time duration of 180 minutes and the teams do not know beforehand when the code will be executed. The ISS orbits the Earth in approximately 90 minutes and, therefore, experiences 2 sunrises and 2 sunsets during 180 minutes. Although night-time photography of bright cities may be possible using the visible camera Izzy, that is not the case for clouds, especially when one needs to have a good signal-to-noise ratio to resolve any wave phenomena. Thus, we must ensure that our pictures are taken in the daytime. Furthermore, we must also avoid taking pictures during sunrise and sunset, since it slightly saturates the camera sensor pixels. In terms of instrumentation, the visible camera placed at a window of the Columbus module of the ISS is the Raspberry Pi Camera Module v1, with 5 Megapixels, and a sensor resolution of $2592\times1944$, with a pixel size of $1.4\,\mathrm{\mu m}$, and a focal length of $3.6\,\mathrm{mm}$. Using the sensor dimensions and the focal length, one can estimate the horizontal and vertical FOV of the camera through equation~(\ref{eq:fov}), which gives approximately $53.5\,\mathrm{degrees}$ and $41.4\,\mathrm{degrees}$, respectively.

The experimental data that can be saved has a limit of 3 Gigabytes (GB) in size. Therefore, one must make a trade-off between the amount of data (e.g., number of images) saved, and its size (i.e., quality and image resolution). By analyzing images taken in previous years, we decided to save images with the maximum resolution available ($2592\times1944$) to better resolve the gravity waves ripples over clouds. The image format used was JPEG (24 bits) with compression at 100\%. On one hand, we lose a small amount of image quality, but on the other hand, we have the advantage of acquiring more images and increase our chances of detecting gravity waves in clouds. Information about events and outcomes during the experiment is recorded and saved in a \textit{log} file.

\subsection{Algorithm}

\begin{figure}
\centering
\includegraphics[width=0.49\textwidth]{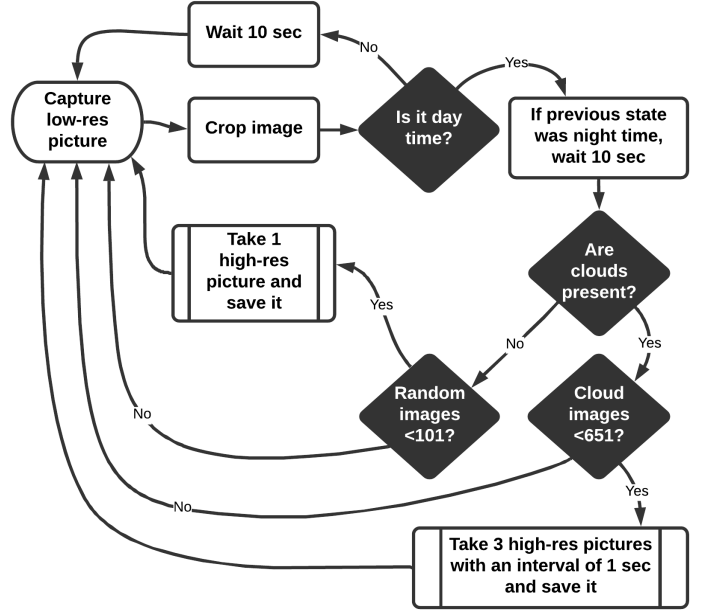}
\caption{\label{fig:flow} Simplified algorithm flowchart of the code developed. The loop starts at "Capture low-res picture".}
\end{figure}

We will now discuss the custom-made code we developed to take pictures of Earth from the ISS. The ultimate goal is to take a maximum number of 720 photographs of clouds in the daytime. A brief version of this code is described in Algorithm~\ref{alg:general} and a simple algorithm flowchart is represented in Fig.~\ref{fig:flow}. 

The code consists mainly of one loop that will run continuously as long as two conditions are satisfied: (i) the total number of recorded pictures is smaller than 720 and (ii) the time passed since the beginning of the loop is less than 177 minutes. The first condition prevents the program from saving data larger than 3 GB, while the second condition ensures that we do not exceed the time limit (180 minutes). The first step in the loop is to take a low-resolution ($640\times480$) picture, for faster processing, and convert it into an image array (Python NumPy array) with shape $640\times480\times3$, where the number 3 refers to the red, green, and blue (RGB) elements of each pixel. The second step consists of cropping the image array to remove the borders of the ISS window where the camera is placed. After this transformation, the image array ends up with a shape of $288\times288\times3$. The contrast of the image is then increased for two purposes: (i) to raise the contrast between dark and light colors and (ii) to raise the contrast between white color and other colors, namely, blue. 
\begin{algorithm}[t]
\caption{Simplified description of the code used aboard the ISS.}\label{alg:general}
\begin{algorithmic}[1]
\State PhotoCount $\leftarrow$ 0 \Comment{\textcolor{blue}{Counter for total images saved.}}
\State RandomCount $\leftarrow$ 0 \Comment{\textcolor{blue}{Counter for random images saved.}}
\State CloudCount $\leftarrow$ 0 \Comment{\textcolor{blue}{Counter for cloud images saved when function \Call{Clouds}{} returns True.}}
\State MaxRandom $\leftarrow$ 101 \Comment{\textcolor{blue}{Maximum number of photos taken when \Call{Clouds}{} function returns False.}}
\State MaxCloud $\leftarrow$ 651 \Comment{\textcolor{blue}{Maximum number of photos taken when function \Call{Clouds}{} returns True.}}

\While {PhotoCount $<$ 720 \AND current time $<$ 177 min} 
    \State EmptyArray $\leftarrow$ Create empty array
    \State LowResImage $\leftarrow$ \Call{Capture}{EmptyArray} \Comment{\textcolor{blue}{Take low-resolution photo and save it into EmptyArray.}}
    \State CropImage $\leftarrow$ \Call{Crop}{LowResImage} \Comment{\textcolor{blue}{Crop the LowResImage array to remove the ISS window borders.}}
    \State TestImage $\leftarrow$ \Call{Constrast}{CropImage} \Comment{\textcolor{blue}{Increase contrast of the CropImage array.}}    
    \If{\Call{DayTime}{TestImage}}
        \If{previously was night-time}
            \State{Wait 10 seconds} \Comment{\textcolor{blue}{Avoid sunshine}}
        \EndIf      
        \If{\Call{Clouds}{TestImage} \AND CloudCount$<$MaxCloud}
        
            \For{$i=0$ to $2$ \textbf{step} 1} \Comment{\textcolor{blue}{Take sequence of 3 photos.}}
            \State Capture high-resolution photo and save
            \State CloudCount$\leftarrow$CloudCount+1
            \State Wait 1 second
            \EndFor
        \Else
            \If{RandomCount $<$ MaxRandom}
                \State Capture high-resolution photo and save
                \State RandomCount$\leftarrow$RandomCount+1
                \State Wait 2 seconds      
            \EndIf          
        \EndIf
    \Else
        \State{Wait 10 seconds.} 
    \EndIf
\EndWhile
\end{algorithmic}
\end{algorithm}
Finally, the modified image array is ready to be used in the input of the two main functions of the algorithm: $\textsc{DayTime}$ and $\textsc{Clouds}$, which are described in Algorithms~\ref{alg:daytime} and~\ref{alg:clouds}, respectively, in~\ref{sec:B}.

To determine if the picture was taken during the daytime or not, the function $\textsc{DayTime}$  is used (see Algorithm~\ref{alg:daytime}). If the picture was taken during night-time, the program waits 10 seconds and returns to the beginning of the loop. On the other hand, if the picture is taken during the daytime, it advances to the next step and checks if the previous outcome was different, i.e., if the last outcome of the $\textsc{DayTime}$ function was false (night-time). If true, then this can possibly be a sunrise and, therefore, the program waits 10 seconds. Otherwise, it advances to the next step, which is to evaluate if clouds are present in the image. Note that, at this point, we are still using the modified image array with a shape of $288\times288\times3$. 

To maximize the probability of finding clouds in the region the camera is photographing, the function $\textsc{Clouds}$ (see Algorithm~\ref{alg:clouds}) is used.  If this function return True, meaning that clouds are probably present, the algorithm takes three pictures, separated by approximately one second, and updates the counters for the number of cloud images (according to the function $\textsc{Clouds}$) and the number of total images that are saved in the system memory. If clouds are not detected according to the function $\textsc{Clouds}$, the program has an extra feature in the extreme case that this function always returns $\textit{false}$ during the mission. The idea is to save "random" pictures during the daytime up to a certain limit. If the function $\textsc{Clouds}$ returns $\textit{false}$, the program checks if the number of saved "random" images is less than a given maximum amount defined initially (in our case, 101). If true, it takes 1 picture and waits 2 seconds. If we already have the maximum amount of "random" pictures, the program goes to the beginning of the loop. 

In the best-case scenario, if the \textsc{DayTime} function works properly, no night-time pictures will be saved and, therefore, we will have 720 pictures taken in the daytime. In terms of cloud pictures, we will have at most 651 "cloud" pictures (i.e., when function \textsc{Clouds} returns true) and, consequently, 69 "random" pictures.

\section{Results}
\label{sec:Results}

\subsection{Overview}
\begin{figure}
\centering
\includegraphics[width=0.49\textwidth]{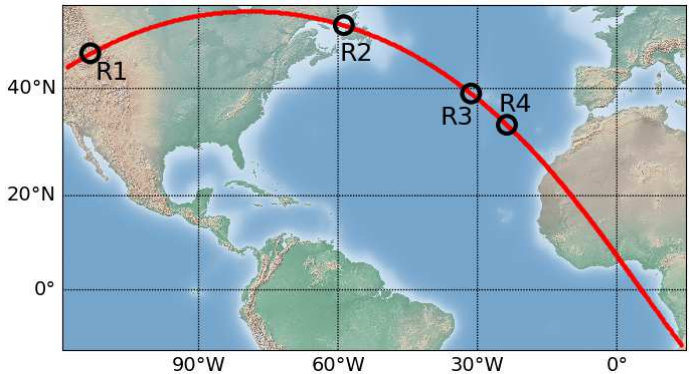}
\caption{\label{fig:trajectory} (Color online) Trajectory of the ISS where pictures were taken. The red solid line is composed of 720 dots corresponding to the coordinates of the ISS where each picture was taken. The dark circles correspond to regions (R1, R2, R3, and R4) where gravity waves were observed and analyzed.}
\end{figure}

The program was executed on the Raspberry Pi computer Izzy at the ISS on April 27, 2020, from 12:06 p.m. to 1:15 p.m. (GMT). According to our \textit{log} file, our program started during night-time at coordinates (-48º18'16.1" S, 139º02'50.4" E), but after approximately 34 minutes, daytime was detected. The total number of images saved was 720, none of which was taken at night-time, which means that the \textsc{DayTime} function accomplished its main goal. The number of "cloud" pictures saved was 651, while the number of "random" pictures was 69. The majority of pictures were taken in the northern hemisphere. The trajectory of the ISS where the 720 pictures were taken is represented in Fig.~\ref{fig:trajectory}.
Four regions of interest containing atmospheric gravity waves were selected. Region 1 (R1), with coordinates (45º 19' 18.0" N, 114º 0' 32.8" W), is near the Jumbo mountain, United States of America. Region 2 (R2), with coordinates (49º 43' 59.1" N, 58º 55' 33.3" W), is in the Gulf of Saint Lawrence, Canada. Region 3 is in the Atlantic ocean, with coordinates (39º 8' 50.9" N, 31º 22' 12.4" W), near the island of Flores, the Azores, Portugal. Finally, region 4 (R4) is also in the Atlantic ocean, with coordinates (33º 46' 29.6" N, 23º 45' 57.0" W), in an area between the islands of the Azores and Madeira, Portugal. The original pictures taken of these regions during the experiment can be found in the supplementary material in Figs.~S1-4. 

To compute the horizontal wavelengths of gravity waves in an image, one has to know the spatial resolution $s'$ of the clouds where "ripples" generated by gravity waves are present. As discussed previously in Section~\ref{sec:spatial_resolution} and stated in equation~(\ref{eq:SRC}), we first need to estimate the spatial resolution $s$ of the surface and then the clouds' altitude $h_\mathrm{c}$ to calculate $s'$. We next discuss the estimation of this two quantities and, finally, the computation of the horizontal wavelength of atmospheric gravity waves present in clouds. 
\begin{figure*}
\centering
\includegraphics[width=0.70\textwidth]{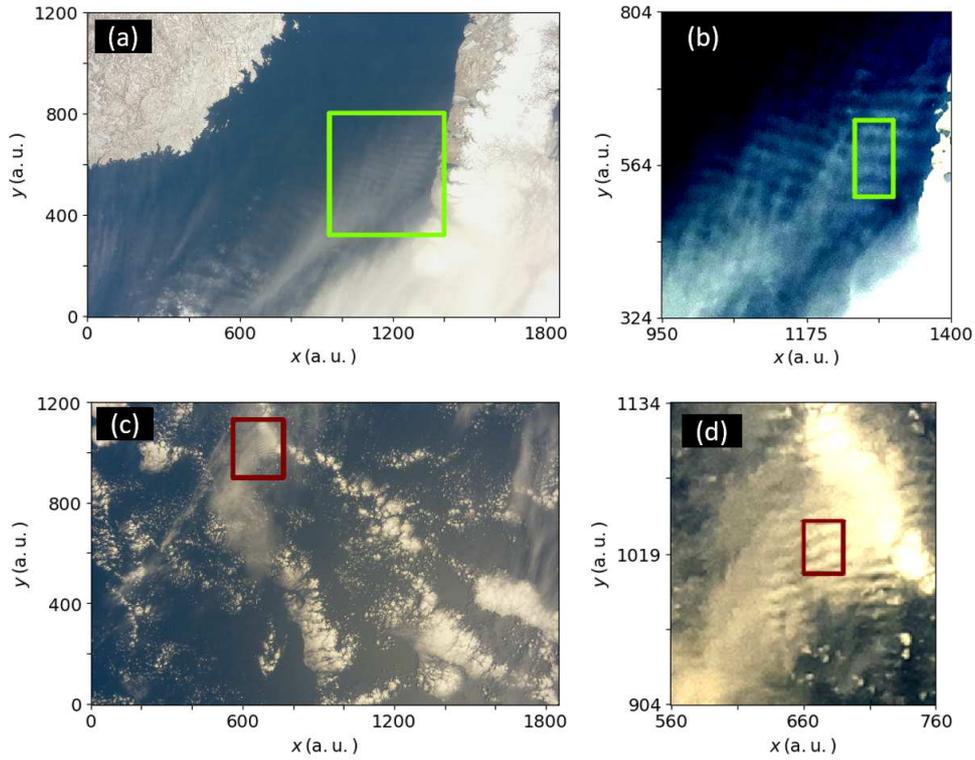}
\caption{\label{fig:region} (Color online) Regions 2 and 4. (a) Picture of region 2 at the Gulf of Saint Lawrence, Canada. (b) Selected domain [green (light) rectangle in (a)] with increased contrast where atmospheric gravity waves are observed. The area used for the Fourier analysis is delimited by the green (light) rectangle. (c) Picture of region 4, situated between the Azores, and Madeira, Portugal. (d) Selected domain [red (dark) rectangle in (c)]. The area used for the Fourier analysis is delimited by the red (dark) rectangle.}
\end{figure*}
\begin{figure*}
\centering
\includegraphics[width=0.68\textwidth]{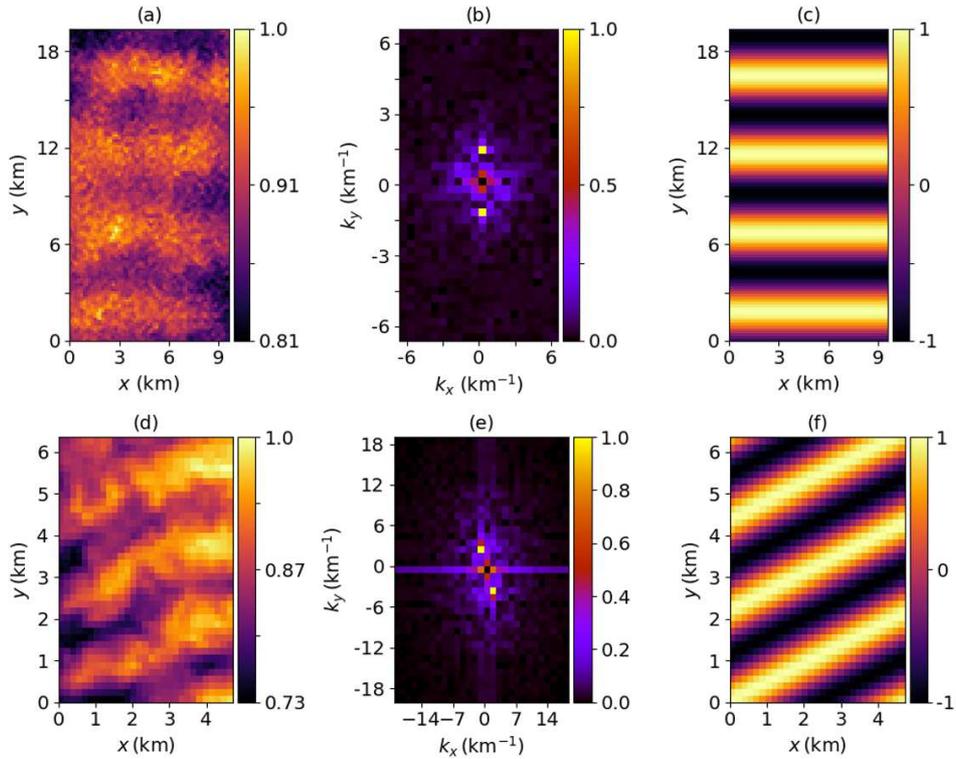}
\caption{\label{fig:fourier} (Color online) Fourier analysis of regions 2 and 4. (a), (d) Selected areas of regions 2 and 4 [see Figs.~\ref{fig:region}(b) and~\ref{fig:region}(d)]. Both images were converted to grayscale. (b), (e) Magnitude of the two-dimensional Fourier transform of (a) and (d), respectively. The magnitude of the central spatial frequency, $\mathbf{k}=(0,0)$, was set to zero for a better visualization. (c), (f) Magnitude of the inverse two-dimensional Fourier transform of (b) and (e), respectively, of the selected wavenumber.}
\end{figure*}

\subsection{Estimation of the ground spatial resolution}

\label{subsec:res_s}
To estimate $s$ we used the two methods mentioned in Section~\ref{subsec:s}. For the first method, we proceeded as follows. For each region (R1-4), we measured distances of land features in units of pixels (for a given image) and compared them with the same distances obtained using Google Earth. For example, in Fig.~\ref{fig:region}(a) we measured the distances on the coast of the Gulf of Saint Lawrence. In cases where the image does not contain land (i.e., image from an ocean region), we used images taken nearby that have land. For example, in region 4 we used images taken nearby of the Azores islands, namely, the island of Pico, to measure distances. Using this analysis, we obtained values for $s$ in the range of 161 to 163 meters per pixel. Note that, since the horizontal wavelengths that are analyzed are smaller than $1000\,\mathrm{km}$, we can neglect the effect of the Earth's curvature~\cite{hines1968}.  For the second method, we simply used the ISS altitude, the camera's focal length and pixel size discussed in Section~\ref{subsec:overview}. The spatial resolution $s$ calculated through equation~(\ref{eq:s2}) gives $s=161\,\mathrm{m}$, which is within the range of values obtained when using the Google Earth method. 

\subsection{Estimation of the clouds' spatial resolution}

\begin{table}[t]
\protect\caption{Impact of the clouds' and ISS altitudes, $h_\mathrm{c}$ and $h_\mathrm{s}$, in the clouds' spatial resolution
$s'$. \label{tab:sl}}
\centering
\begin{tabular}{|c|c|c|c|}
\hline 
$s\,(\mathrm{m})$ & $h_{\mathrm{c}}\,(\mathrm{km})$ & $h_{\mathrm{s}}\,(\mathrm{km})$ & $s'\,\mathrm{(m)}$\tabularnewline
\hline 
\hline 
\multirow{4}{*}{161} & 0.1 & \multirow{2}{*}{385} & 161\tabularnewline
\cline{2-2} \cline{4-4} 
 & 15 &  & 155\tabularnewline
\cline{2-4} 
 & 0.1 & \multirow{2}{*}{445} & 161\tabularnewline
\cline{2-2} \cline{4-4} 
 & 15 &  & 156\tabularnewline
\hline 
\end{tabular}
\end{table}
After estimating $s$, we are left with estimating the clouds' altitude $h_\mathrm{c}$ to retrieve the spatial resolution $s'$ of the clouds which contain oscillation patterns. To this end, we identified the cloud genera where gravity waves are present so that we can determine their range of altitudes (see Table~\ref{tab:genera}). Clouds of R2 and R4, where the effects of gravity waves are present, were classified as being of type Cirrus, while those of R1 was classified as being of type Cirrostratus. For R3, however, it was unclear if their type was Altostratus or Cirrostratus and, therefore, we have chosen a wider interval that takes into account both types.  Considering  $h_\mathrm{c}$ as the average value of the cloud altitude interval and assuming an average altitude of the ISS of $h_\mathrm{s}=415\,\mathrm{km}$, the spatial resolution $s'$ of clouds is readily obtained using equation~(\ref{eq:SRC}). If we consider a maximum cloud altitude of $h_\mathrm{c}=15\,\mathrm{km}$, it will change the cloud spatial resolution by approximately $4\%$. Table~\ref{tab:sl} shows the impact of the extreme values of $h_\mathrm{c}$ and $h_\mathrm{s}$ in the clouds' spatial resolution $s'$. The error associated with $s'$ is discussed in~\ref{sec:A}.

\subsection{Computation of the horizontal wavelength of gravity waves}

\begin{table*}[t!]
\caption{Results for the atmospheric gravity waves horizontal wavelength. NW, NE, SW, and SE denote the northwest, northeast, southwest and southeast, respectively.}
\label{tab:results}
\begin{centering}
\begin{tabular}{|c|>{\centering}p{0.25\linewidth}|c|c|c|c|}
\hline 
Region & Cloud Genera & $h_{\mathrm{c}}$(km) & $s'\,(\mathrm{m})$ & $\lambda_{h}\,(\mathrm{km})$  & Direction \tabularnewline
\hline
\hline 
R1 & Cirrostratus & $11\pm4$ &  157 & $1.9\pm0.2$  & NW or SE\tabularnewline
\hline 
R2 & Cirrus & $11\pm4$ & 157 & $4.7\pm0.3$ & NE or SW \tabularnewline
\hline   
R3 & Altostratus / Cirrostratus & $8.5\pm6.5$ & 158 & $1.0\pm0.3$  & NE or SW \tabularnewline
\hline 
R4 & Cirrus & $11\pm4$& 157 & $1.9\pm0.2$  & NE or SW \tabularnewline
\hline 
\end{tabular}
\par\end{centering}
\end{table*}

After estimating $s'$, we selected a domain in each region (R1-4) where atmospheric gravity waves can be observed. Then, we increased the contrast of this domain. The amount of contrast applied varies depending on the region. Figure~\ref{fig:region} shows this procedure for R2 and R4. Figures~\ref{fig:region}(a) and (c) are pictures of regions 2 and 4, respectively, that were cropped to remove the borders of the window. Figures~\ref{fig:region}(b) and (d) are the selected contrast-enhanced domains of Figs.~\ref{fig:region}(a) and~(c), respectively.  The same procedure for the remaining two regions (R1 and R3) can be found in the supplementary material in Fig.~S5.  

For each domain containing gravity waves, we selected a rectangular area where a Fourier analysis is performed to compute the wavenumbers $k_x$ and $k_y$. Assuming that the wavenumbers are real [see equation~(\ref{eq:wavenumber})], we compute the horizontal wavelength $\lambda_h$ using equation~(\ref{eq:hw}). The green and red (light and dark) rectangles of Fig.~\ref{fig:region}(c) and (d), respectively, are the areas chosen for the Fourier analysis. Figure~\ref{fig:fourier} illustrates the several steps of the Fourier analysis for regions 2 and 4. First, the rectangular area is converted to grayscale and normalized [Figs.~\ref{fig:fourier}(a) and~\ref{fig:fourier}(d)]. Then, a two-dimensional Fourier transform is applied [Figs.~\ref{fig:fourier}(b) and~\ref{fig:fourier}(e)]. Here, several high-intensity spatial frequencies may be present that are not due to gravity waves. We select the wavenumber corresponding to the observed pattern and performed an inverse two-dimensional Fourier transform to confirm that it corresponds to the same pattern [Figs.~\ref{fig:fourier}(c) and~\ref{fig:fourier}(f)]. The direct and inverse two-dimensional Fourier transforms were done using the NumPy's fast Fourier transform  algorithm. Figure~S6 of the supplementary material shows the analogous of Fig.~\ref{fig:fourier} for R1 and R3.

Results for the horizontal wavelengths are presented in Table~\ref{tab:results}, along with the estimation of the cloud genera, their average altitude and the possible directions. The horizontal wavelengths analyzed are in the range of $1.0$ to $4.7\,\mathrm{km}$. The uncertainties are obtained by applying the error propagation equation for the horizontal wavelength $\lambda_h$ (see~\ref{sec:A}). Due to the ISS speed, we cannot take pictures of the same place for a period of time longer than a few seconds and, therefore, we can not determine their period. Nevertheless, we can conclude that it is more likely that we are in the presence of small-scale gravity waves since their horizontal wavelengths are much smaller than 10 km. These types of gravity waves may not be internal, i.e., they can be evanescent or ducted waves due to instabilities and restricted to a small range of altitude~\cite{snively2008,simkhada2009}. Since we do not have access to where the wave vector is pointing, we included the two possible directions. We used Google Earth to estimate the North cardinal direction for each region.


\section{Conclusion}
\label{sec:Conclusion}

In this work, we have shown that atmospheric gravity waves can be detected in clouds and partially spatially characterized in terms of their horizontal wavelength using low-cost and educational instrumentation onboard the ISS, namely, a Raspberry Pi computer and a visible camera. We developed a code capable of taking pictures of clouds from the ISS in the context of the Astro Pi Mission Space Lab. We have discussed in detail the algorithm, which tackles several constraints such as night-time, sunrise, and absence of clouds. The code was executed in a Raspberry Pi computer on the Columbus module of the ISS. We were able to detect the presence of atmospheric gravity waves in several pictures  through the oscillatory patterns they produce in clouds. First, we identified the type of clouds (i.e., their genera) where gravity waves were present and estimated their altitude, which enabled us to perform a Fourier analysis. The spatial resolution of the image was sufficient for us to determine the horizontal wavelengths of gravity waves. The resulting values were in the range of $1.0$ to $4.7\,\mathrm{km}$. Even though we do not have access to the period, due to the ISS speed, we can conclude that it is more likely that we are in the presence of small-scale gravity waves. In future work, the algorithm to detect clouds can be improved to better detect clouds and to identify their type, thus increasing the chances of finding gravity waves. With regard to the clouds' altitude determination, one can improve the main algorithm and subsequent analysis by implementing a triangulation method, where one takes advantage of the ISS high speed (too take subsequent pictures) and approximately constant speed.

\section{Authors contribution}
TECM was the mentor and supervisor of this project. TECM, DECGS, CECGS, AAD and JPMM developed the software and performed the data analysis. TECM and TMR performed a literature review and wrote the manuscript. All authors discussed the results, reviewed the manuscript and approved the final manuscript.

\section{Declaration of competing interest}
The authors declare no conflict of interest.

\section{Acknowledgments}
The authors acknowledge the European Space Agency (ESA) and the Raspberry Pi foundation for the opportunity the run this experiment in the International Space Station (ISS). We also thank Isabel Allen for her support throughout this project. The authors are grateful to the referees for their suggestions.

\appendix

\section{Algorithms}
\label{sec:B}

Algorithm~\ref{alg:daytime} receives as input the modified (low-resolution, cropped, and contrast-enhanced) image array (line 10 of Algorithm~\ref{alg:general}) and a threshold value $t_{\text{d}}$. The latter was defined by analyzing images taken from previous editions of the Astro Pi challenge. The \textsc{DayTime} function starts by converting the modified image array into grayscale, ending up with the shape $288\times288\times1$, and then calculates the average value of the $288\times288\times1$ elements. If the average value is larger than the threshold value $t_{\text{d}}$, the function returns true, meaning that the low-resolution picture was taken in the daytime. Otherwise, it returns false. The advantage of this method is the small computation time ($\sim22.6\,\mathrm{ms}$) needed when using the Numpy's $\textsc{mean}$ function.
\begin{algorithm}[b]
\caption{Function used to detect day time.}\label{alg:daytime}
\begin{algorithmic}[1]
\Function{DayTime}{TestImage, $t_{\mathrm{d}}$} 
\State GrayImage $\leftarrow$ \Call{convert2gray}{TestImage} \Comment{\textcolor{blue}{Converts the image to grayscale.}}
\State MeanValue $\leftarrow$ \Call{mean}{GrayImage} \Comment{\textcolor{blue}{Computes the mean value of the pixels, from 0 to 255.}}
\If{MeanValue $>$ $t_{\mathrm{d}}$}
    \State \Return True \Comment{\textcolor{blue}{Day Time}}
\Else
    \State \Return False \Comment{\textcolor{blue}{Night Time}}
\EndIf
\EndFunction
\end{algorithmic}
\end{algorithm}

Algorithm~\ref{alg:clouds} receives as input the modified image array (line 10 of Algorithm~\ref{alg:general}), a threshold value $t_{\text{c}}$, and a threshold RGB color array $(R,G,B)$. The latter defines a threshold color for which the pixel color is considered as being part of a cloud. For example, since the color of clouds varies from gray to white, one can define the threshold color array as $(130,130,130)$, which corresponds to the color light gray. Both quantities were defined by analyzing images taken in previous editions of the Astro Pi challenge. After receiving these three inputs, the function counts the number of pixels for which the color is larger than the threshold color. For this task, NumPy's function $\textsc{count\_nonzero}$ is used. If the total number is larger than the threshold number $t_{\text{c}}$, the function $\textsc{Clouds}$ return true, meaning that clouds are probably present. Function $\textsc{Clouds}$ has some drawbacks since some ocean and land features (e.g., snow) can easily be interpreted as clouds. On the other hand, the small computation time needed ($\sim144\,\mathrm{ms}$) is a great advantage in this case.
\begin{algorithm}[t]
\caption{Function used to detect clouds.}\label{alg:clouds}
\begin{algorithmic}[1]
\Function{Clouds}{TestImage, Tcolor, $t_{\mathrm{c}}$} \Comment{\textcolor{blue}{Returns True if clouds are detected in the image}}
\State CountTcolor $\leftarrow$ \Call{CountColor}{TestImage,Tcolor} \phantom N \phantom N \phantom N  \Comment{\textcolor{blue}{Counts the number of pixels in TestImage for which the RGB components are larger than Tcolor.}}
\If{CountTcolor$>$ $t_{\mathrm{c}}$}
    \State \Return True \Comment{\textcolor{blue}{Clouds detected.}}
\Else
        \State \Return False \Comment{\textcolor{blue}{Clouds not detected.}}
\EndIf
\EndFunction
\end{algorithmic}
\end{algorithm}

\section{Error analysis}
\label{sec:A}
Let $\delta s'$ be the uncertainty associated with $s'$. If we assume that the uncertainties in $h_\mathrm{c}$, $h_\mathrm{s}$ and $s$ are independent and random, and apply the error propagation equation~\cite{Taylor1997} in $s'$ of equation~(\ref{eq:sl}), we find that
\begin{equation}
    \delta s' = \sqrt{\delta s^2-\left(\frac{h_{\text{c}}}{h_{\text{s}}}\delta s\right)^2   +\left(\frac{\delta h_{\text{c}}}{h_{\text{s}}}\right)^2+\left(\frac{h_{\text{c}}}{h_{\text{s}}^2} \,\delta h_{\text{s}}\right)^2}\,,\label{eq:uncer}
\end{equation}
where $\delta s$, $\delta h_{\text{c}}$ and $\delta h_{\text{s}}$ are the uncertainties of the clouds' altitude, of the ISS altitude, and of the surface spatial resolution, respectively. Notice that the second, third, and forth terms inside the square root of equation~(\ref{eq:uncer}) will be very small compared to the first term since $h_\mathrm{s}\gg h_\mathrm{c}$. For instance, in the second term, if we assume $h_\mathrm{c}=15\,\mathrm{km}$ as the  maximum value (see Table~\ref{tab:genera}), then $h_\mathrm{c}^2/h_\mathrm{s}^2\approx0.001$. Therefore, the second term in the square root will be $1000$ times smaller than the first one. The remaining terms will also be small compared to $\delta s$. The main contribution to the uncertainty of $s'$ will come from the uncertainty in $s$. Table~\ref{tab:deltaS} illustrates the impact of the uncertainty in the clouds' altitude $\delta h_\mathrm{c}$ in $\delta s'$.

The horizontal wavelength of gravity waves is computed through the FFT algorithm. Therefore, the spatial resolution $s'$ will ultimately define the spatial frequency resolution $\Delta K_j$ of the FFT algorithm in the $j$-axis through~\cite{Rao2011}
\begin{equation}
    \Delta K_j =\frac{1}{N_j\,s'}\,,
\end{equation}
where $N_j$ is the size of the array in the $j$-axis. In a given image, the minimum sampling wavelength $\lambda_{\min}$ is, according to the Nyquist sampling theorem~\cite{shannon1949},
\begin{equation}
    \lambda_{\min} = 2 s'\,.
\end{equation}
As an approximation, we will consider the uncertainties of $\lambda_x$ and $\lambda_y$, $\delta \lambda_x$ and $\delta \lambda_y$, as $2s'$. Therefore, using the error propagation formula once more for equation~(\ref{eq:hw}), we obtain
\begin{equation}
    \delta \lambda_h = \sqrt{\frac{\left(\lambda_y^3\,\delta\lambda_x\right)^2+\left(\lambda_x^3\,\delta\lambda_y\right)^2}{\left(\lambda_x^2+\lambda_y^2\right)^2}}\,.
\end{equation}
\begin{table}[t]
\protect\caption{Impact of $\delta s$ and $\delta h_\mathrm{c}$ on $\delta s'$. Notice that $\delta s'/s'$ is only affected by the change in $\delta s$. \label{tab:deltaS}}
\centering
\begin{tabular}{|c|c|c|c|c|c|c|c|c|}
\hline 
$s\,(\mathrm{m})$ & \multicolumn{8}{c|}{161}\tabularnewline
\hline 
$h_{\mathrm{s}}\,(\mathrm{km})$ & \multicolumn{8}{c|}{415}\tabularnewline
\hline 
$\delta h_{\mathrm{s}}\,(\mathrm{km})$ & \multicolumn{8}{c|}{30}\tabularnewline
\hline 
$\delta s\,(\mathrm{m})$ & \multicolumn{4}{c|}{8} & \multicolumn{4}{c|}{16}\tabularnewline
\hline 
$\delta h_{\mathrm{c}}\,(\mathrm{km})$ & \multicolumn{2}{c|}{0.1} & \multicolumn{2}{c|}{15} & \multicolumn{2}{c|}{0.1} & \multicolumn{2}{c|}{15}\tabularnewline
\hline 
$h_{\mathrm{c}}\,(\mathrm{km})$ & 0.1 & 15 & 0.1 & 15 & 0.1 & 15 & 0.1 & 15\tabularnewline
\hline 
$\delta s'/s'$ & \multicolumn{4}{c|}{5\%} & \multicolumn{4}{c|}{10\%}\tabularnewline
\hline 
\noalign{\vskip2pt}
\end{tabular}
\end{table}

\bibliography{mybibfile}

\end{document}